\documentclass[journal,12pt,onecolumn,draftclsnofoot,]{IEEEtran}
\usepackage{cite} % Tidies up citation numbers.
\usepackage{url} % Provides better formatting of URLs.
\usepackage[utf8]{inputenc} % Allows Turkish characters.
\usepackage{booktabs} % Allows the use of \toprule, \midrule and \bottomrule in tables for horizontal lines
\usepackage{graphicx}
\usepackage{amssymb}
\usepackage{amsmath}
\usepackage{hyperref}
\usepackage{todonotes}
\usepackage{listings}
\hypersetup{
    colorlinks=true,
    linkcolor=blue,
    filecolor=magenta,      
    urlcolor=cyan,
}

\begin{document}
%\begin{titlepage}
\title{EventDetectR- An Open-Source Event Detection System}


\author{Sowmya Chandrasekaran, Margarita Rebolledo, Thomas Bartz-Beielstein \\
Institute for Data Science, Engineering, and Analytics, TH K\"oln \\ Steinmüllerallee 1, 51643 Gummersbach, Germany\\
}
%\date{25/4/15}

\maketitle
\tableofcontents
\listoffigures
\listoftables
%\end{titlepage}

\IEEEpeerreviewmaketitle
\begin{abstract}
\href{https://cran.r-project.org/web/packages/EventDetectR/index.html}{EventDetectR}: An efficient Event Detection System (EDS) capable of detecting unexpected water quality conditions. This approach uses multiple algorithms to model the relationship between various multivariate water quality signals. Then the residuals of the models were utilized in constructing the event detection algorithm, which provides a continuous measure of the probability of an event at every time step. The proposed framework was tested for water contamination events with industrial data from automated water quality sensors. The results showed that the framework is reliable with better performance and is highly suitable for event detection.
\end{abstract}

\section{Introduction}
The quality of drinking water is a major concern as it directly affects public health and safety. Hence the protection of the drinking water supply and distribution systems against accidental or intentional contamination is crucial. The current technological advancements have made more and more affordable online sensors available. The recent rapid breakthroughs in statistical analysis and more low-cost online sensors have enabled the development of Event Detection System (EDS).
\\
EDS is required to analyze the huge volume of water quality data from various monitoring stations, to detect unexpected water quality conditions, and to alarm the operator to handle these situations. Such an EDS can be combined with static set-point or threshold alarms. A clear difference between set-point alarms and EDS is that set-point alarms are simply triggered when a water quality signal breaches the specified control limits. In contrast, EDS is designed to identify the abnormal behavior of the water quality.\\

 In \cite{Stor11a}, an extensive study about the recent advances in the online drinking water monitoring systems is carried out. Their report conveys that mostly commercially available tools provide reliable performance. \\ 

 Some of the commercial tools include OptiEDS-OptiWater~\cite{Ostf04a}, GuardianBlue Event Detection System by HACH \cite{hach00a}, ana::tool an event detection software \cite{scan00a}. Except for commercial software, there are only a few event detection systems like CANARY~\cite{Murr10a}. Therefore, we intend to develop an open-source Event Detection System that is efficient and is capable of detecting unexpected water quality conditions. The proposed framework can be easily extended as a real-time monitoring system.\\

 The rest of this report is structured as follows: Section~\ref{sec:1} presents an introduction and background to the research project. Then Section~\ref{sec:2} clearly outlines the existing state of the art event detection methods. Thereafter, Section~\ref{sec:3} explains the proposed event detection approach. It also focuses on training the event detection framework and identifying the appropriate parameter settings. Then in Section~\ref{sec:4}, the suitable performance metrics are discussed. In Section~\ref{sec:5} a case study is presented. Finally, a summary and an outlook is presented in Section~\ref{sec:6}.  
\\
\section{Background of the Research Project}\label{sec:1}
 With the current advancements in measurement technology, affordable online sensors are being installed in the drinking water production and distribution systems. They provide a huge amount of real data for analysis. However, a huge fraction of this data is not effectively used by the operators to gain a better understanding of the current status of their system. This brings in the necessity to devise an EDS that could automatically utilize the available data and infer vital information about the water quality. This can then aid the operators with easy proactive maintenance and can help in providing safe drinking water to the public.\\
 
 \subsection{Open Water Open Source(OWOS) Project}
The OWOS project research focuses mainly on the drinking water quality in Germany, taking into account the effects of climate change, energy efficiency, and protection against environmental disasters and terrorist attacks. As an outcome of this project, the \href{https://cran.r-project.org/web/packages/EventDetectR/index.html}{EventDetectR} and \href{https://cran.r-project.org/web/packages/EventDetectGUI/index.html}{EventDetectGUI} Packages are developed as an open-source project. It focuses on the goal of monitoring the quality of drinking water in Germany. \\

The major goals of the project are:
\begin{itemize}
    \item How can we ensure drinking water quality with increasing extreme weather conditions?
    \item How can drinking water suppliers control their processes in an energy and resource-efficient way?
    \item How can consumers be protected from the effects of environmental disasters?
  \end{itemize}  
  
  In order to handle the above-discussed goals, the project research interests are: 
  
  \begin{itemize}
  \item Process for the determination of trend analysis, long-term water supply forecasting
 \item Area-specific adaptation of the monitoring networks and monitoring programs, thus enabling reliable estimates of changes in water quality
 \item Methods for intelligent resource management
 \item Network monitoring to adapt the individual drinking water process
\item Optimization of the sensor placement in order to better detect disturbances, drift, and failures
\item Optimization of the drinking water network
\item Simulation of different contamination scenarios
\item Development of an event detection system
\item Providing an event simulator
\item Development of an alarm system
\end{itemize} 
\subsection{Drinking Water Supply and Distribution Systems} 
\begin{figure}[h]
\centering
\includegraphics[width=1\textwidth]{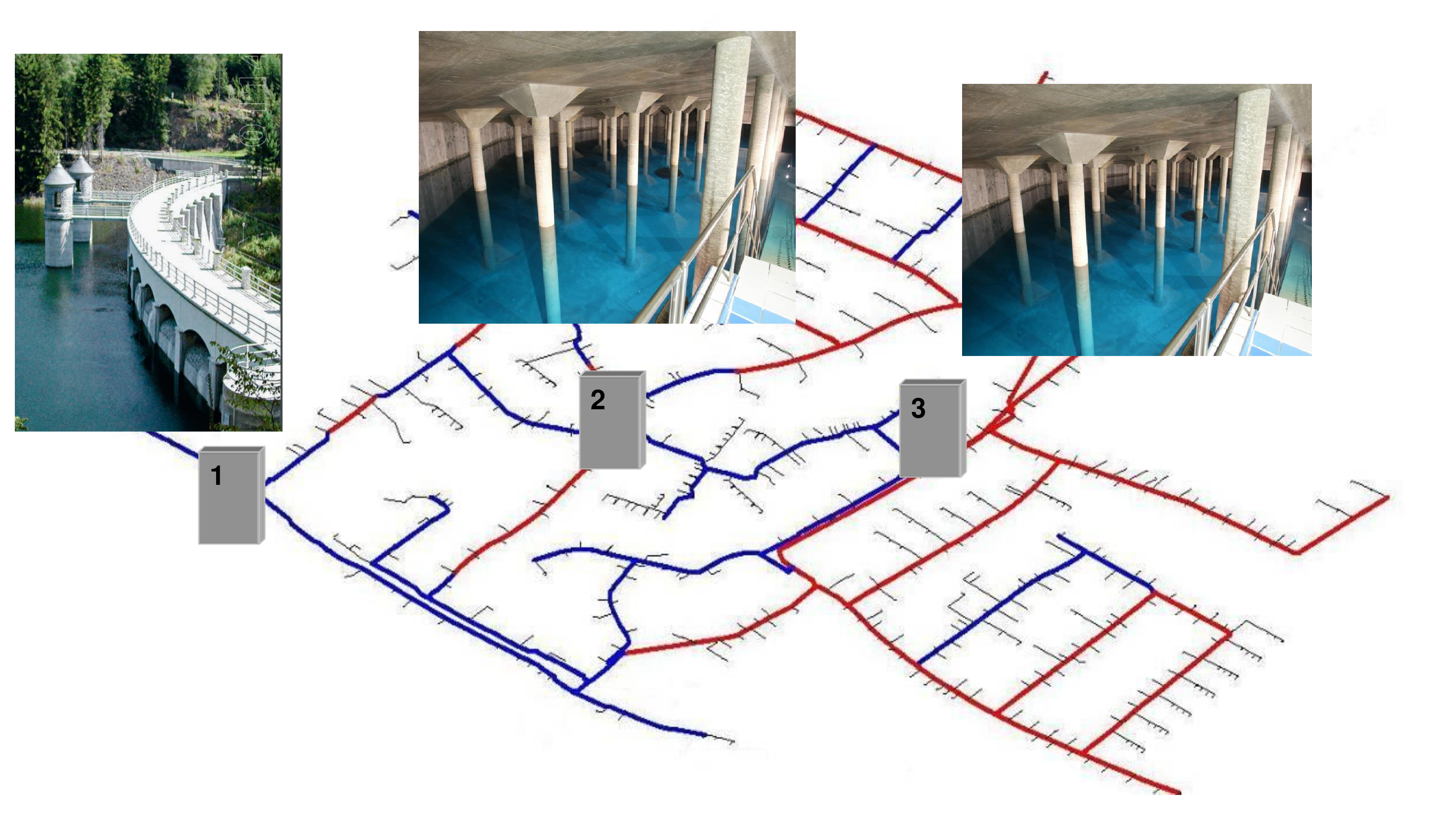}
\caption[Sensors installed at Monitoring Stations]{Sensors installed at various Monitoring Station in the Water Distribution Network}
\label{fig:sensor}
\end{figure}

The drinking water supply and distribution systems generally include storage tanks, pipes, pumps, valves, reservoirs, meters, fittings, and other hydraulic appliances. Their network distribution ranges from source water well to a water treatment plant and traverses through various complex channels until it reaches the end customers' tap. Along with the system, the possibility of contamination injection into the distribution system either intentionally or unintentionally is significant. Fool-proof physical securing of the complete distribution network is not realizable. A reliable alternative is the installation of  monitoring stations to measure the water quality with the help of sensors located at each monitoring station, as shown in Figure~\ref{fig:sensor}. 

\subsection{What is an Outlier, Event, and a Baseline Change?}
\begin{figure}[h]
\centering
\includegraphics[width=1\textwidth]{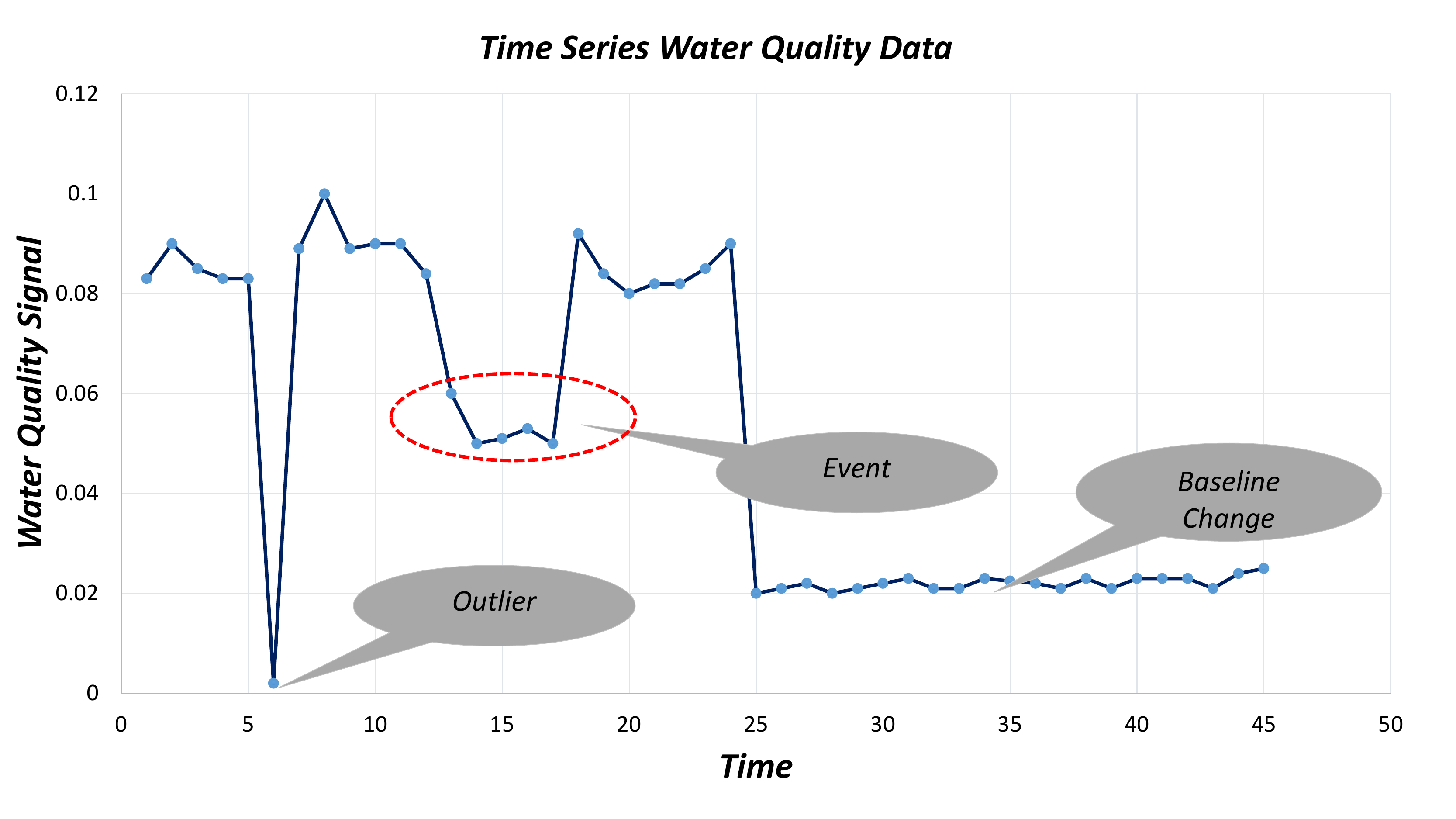}
\caption{Representation of an Outlier, Event and a Baseline Change}
\label{fig:event}
\end{figure}

An event can be defined as a significant deviation of one or more water quality parameters from their normal behavior for more than a specific period. As it can be seen in Figure~\ref{fig:event}, a significant deviation of a water quality parameter at a single time step is an outlier. This is mostly due to a short-term measurement problem (e.g., air bubbles in the line) and should not lead to an alarm message. When multiple outliers occur over a specific period, then it is defined as an event. An event is an actual deviation from normal behavior that is to be reported. The deviation must last for a certain period. The reported deviation from normal behavior does not necessarily have to represent an issue in the quality of the water. Changes in operation or maintenance should also be reported as a deviation. After a change in certain operating parameters, a baseline change is said to have occurred. A baseline change is a new normal behavior. In such a case, no more events should be reported. These time limits for events and baseline changes differ from utility to utility and are user-defined. 

\subsection{Challenges faced by an Event Detection System} 
 The main objective of an EDS is to capture the unexpected behavior as quick as possible and alert the operator. The major challenge is the rate of false alarms, which should be kept at a minimum. There is always a trade-off required between accuracy, false alarm, and quick signaling of unexpected behavior. The following challenges may lead to a false alarm:
\begin{itemize}
\item Differentiating a contamination event from a naturally deviating water quality characteristic
\item Accurate testing of an EDS is very complicated
\item Malfunctioning of a sensor due to various external or internal factors. 
\item Data transmission issues which may lead to incomplete data
 \end{itemize}
 
\section{Background on Event Detection Algorithms in Drinking Water Systems}\label{sec:2}
The use of data mining algorithms for event detection in drinking water quality control can be traced back to \cite{Hart07a} and \cite{Klis06a}. A Linear Prediction Coefficient Filter (LPCF) algorithm was introduced for water quality event detection. It predicts the water quality at a future time step using recent observations. It uses an Auto Regressive (AR) model and then evaluates the residuals between predicted and measured water quality values. When the threshold value for the residuals is exceeded, events are triggered. Also, Multivariate Nearest Neighbour (MVNN) algorithm is used to classify the current observation as normal or anomalous by calculating multivariate Euclidean distance was developed.\\

In \cite{Smet06a}, the use of the Statistical Process Control (SPC) techniques that deal with turbidity data is demonstrated. The issue of auto-correlation and its effect on the performance of SPC was extensively discussed. The usage of ARIMA model residuals for the construction of SPC charts was recommended.\\ 

CANARY  was introduced as an online streaming event detection tool based on MATLAB that can process data from multiple sensor locations\cite{Murr10a}. Algorithms in CANARY include Set Point Proximity Exponential (SPPE) and Set Point Proximity Beta (SPPB) algorithms, used for the proximity of the event calculation. It also provides options to combine either of the set-point algorithms with LPCF or MVNN using Consensus MAXimum (CMAX) or Consensus AVErage (CAVE) algorithms. The CMAX algorithm takes the maximum probability from both algorithms that are combined, while CAVE algorithm averages the results of both. One existing issue with CANARY is altering the threshold dramatically changes the performance of the event detection and the ability to detect anomalous water quality signals.\\

 The use of ANN for classification of water quality signals into normal and anomalous classes is performed in \cite{Pere12a}. Here,  ANN is used to detect possible outliers. Then the probability of an event is updated using Bayes' rule. The probabilities are used for identifying anomalous behavior. An improvement for \cite{Pere12a} was proposed in \cite{Arad13a}, which involved dynamic threshold limits. In \cite{Hou13a}, an event detection technique based on multi-sensor fusion using an extended Dempster Shafer (DS) method was proposed. Initially, an AR model is used to predict future time step values, then probabilities are assigned to the residuals obtained at each time step. Then DS fusion method searches for anomalous probabilities of the residuals. \\

 A classification-based approach for event detection is carried out in \cite{Olik14a}, where support vector machines were employed to detect outliers, and sequential analysis is carried out to classify events. 

In \cite{Liu15a}, clustering based analysis is implemented to classify the type of contaminant present in the water systems. The Mahalanobis distance of the water quality vectors classifies the type of contaminant in the water distribution systems based on the similarities of sensors responses to predefined classes.\\

A cloud based EDS is presented in \cite{Kuhn15a}. 
In \cite{Puig17a}, various approaches for monitoring drinking water in real time are presented. Related studies were also recently carried out in  \cite{Leow17a}, \cite{Bila18a}, \cite{Muha18a}. 

\section{\href{https://cran.r-project.org/web/packages/EventDetectR/index.html}{EventDetectR}- R Package}\label{sec:3}
The EventDetectR Framework detects events in a multi-dimensional time-series data. For each incoming new water quality parameter, the framework has to classify whether this multivariate measurement is a contamination event or not. This is done through various steps using model residuals, thresholds, and the probability of an event.\\

\begin{figure}[h]
\centering
\includegraphics[width=1\textwidth]{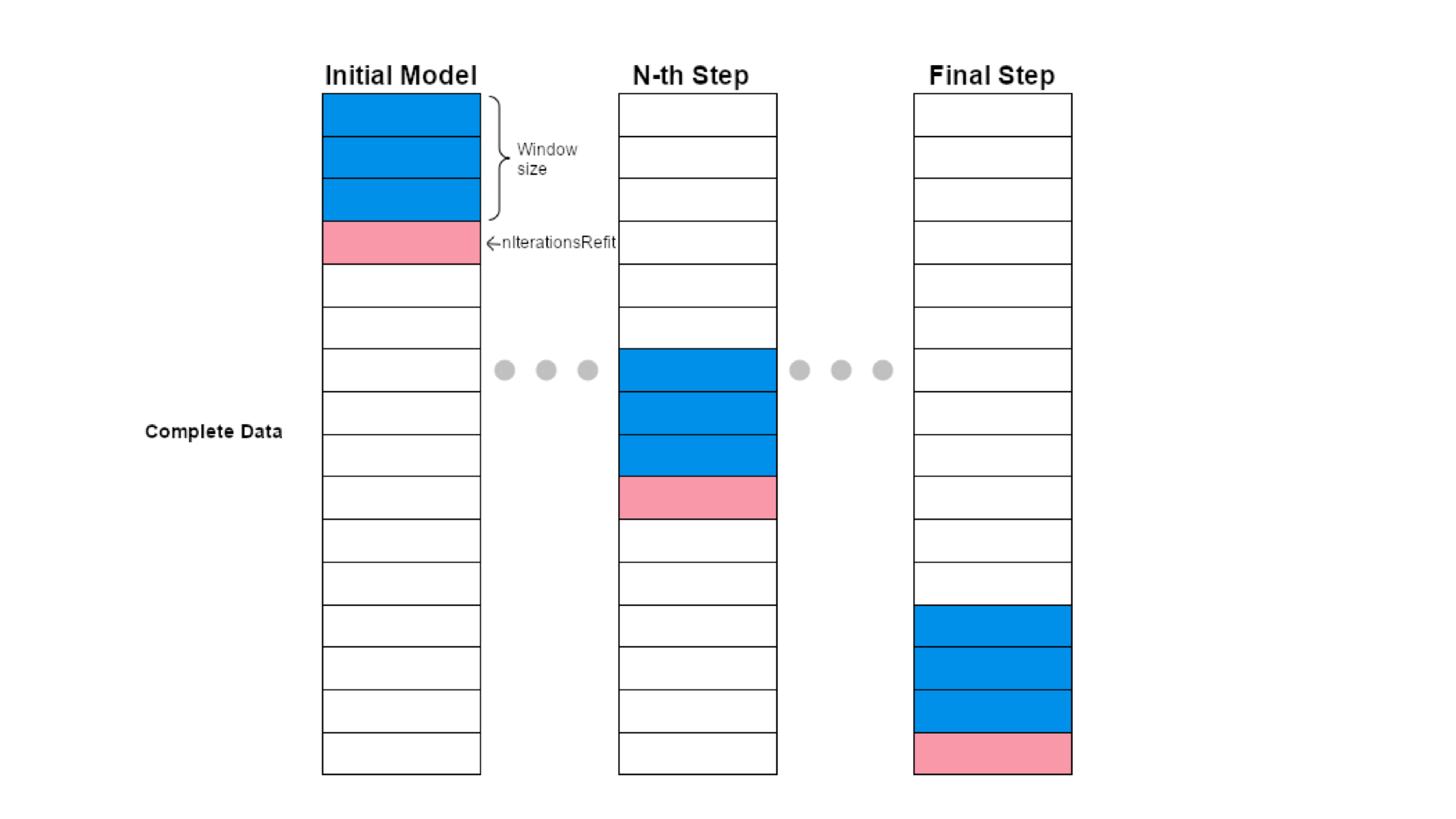}
\caption[EventDetectR-Process Flow]{EventDetectR-Process Flow}
\label{fig:e2}
\end{figure}

\begin{figure}[h]
\centering
\includegraphics[width=1\textwidth]{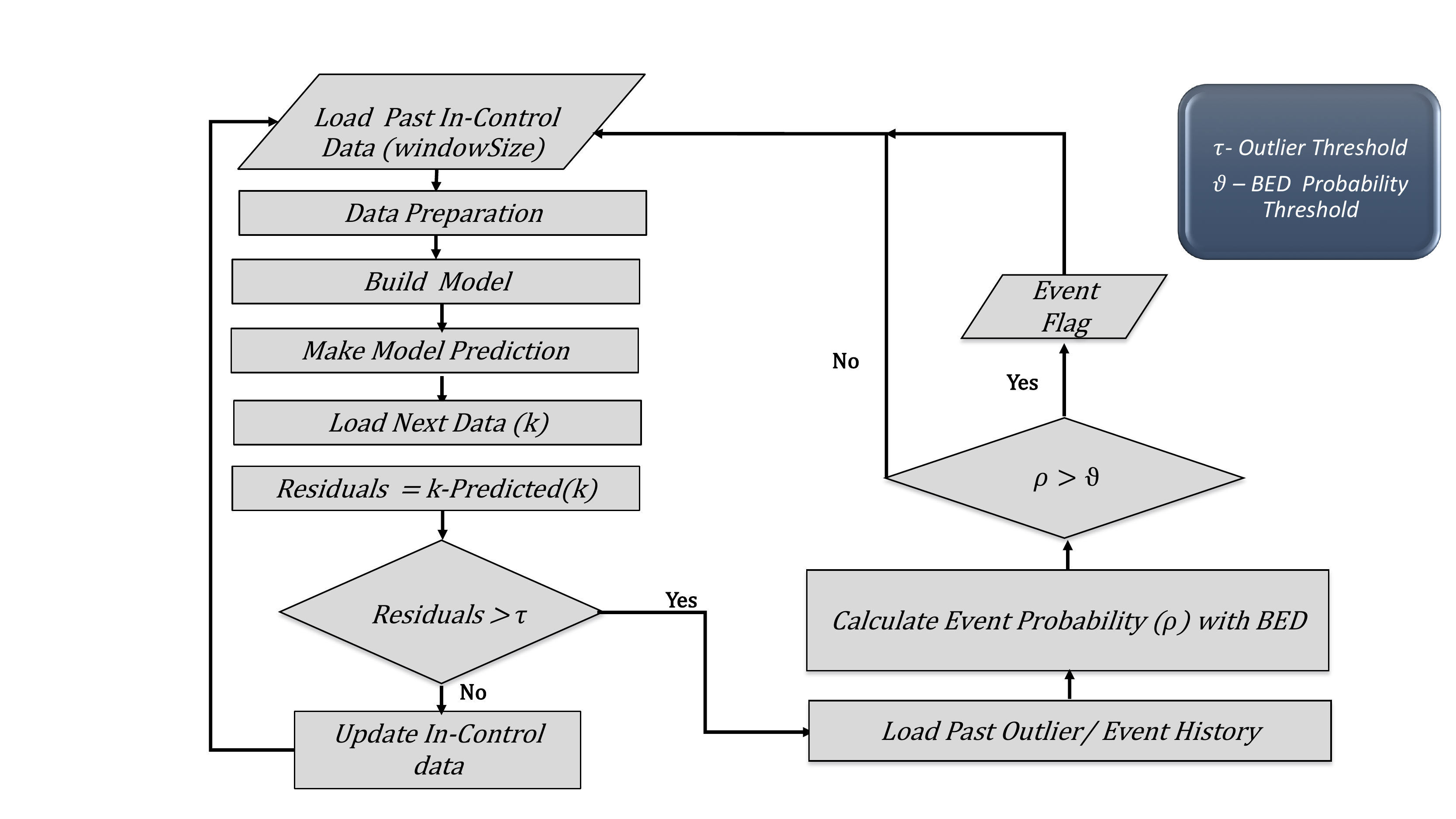}
\caption[EventDetectR-Methodology Flowchart]{EventDetectR- Methodology Flowchart}
\label{fig:e1}
\end{figure}

The event detection process flow can be visualized in Figure \ref{fig:e2} and Flowchart \ref{fig:e1}. The $windowSize$ is the sliding window, which represents the number of past observations used to build the model at each step. At each classification iteration, a window of $windowSize$ datapoints (shown in blue) enters into the data preparation phase and the older ones exit in a first in first out fashion. Then, a suitable model is fitted and used to predict the next $nIterationsRefit$ data into the future. The real data (shown in red) is then compared to the prediction and the difference between them is obtained as residuals. These residuals, together with the specified classification thresholds, are used to decide which data is considered an outlier and which is considered as normal behavior. BED is then employed to determine the probability of an event. After the classification, the window is moved by $nIterationsRefit$ rows in the data, and the procedure is repeated until the end of the time-series is reached, and thus all elements are classified.

\subsection{The Event Detection Phases}
\begin{figure}[h]
\centering
\includegraphics[width=1\textwidth]{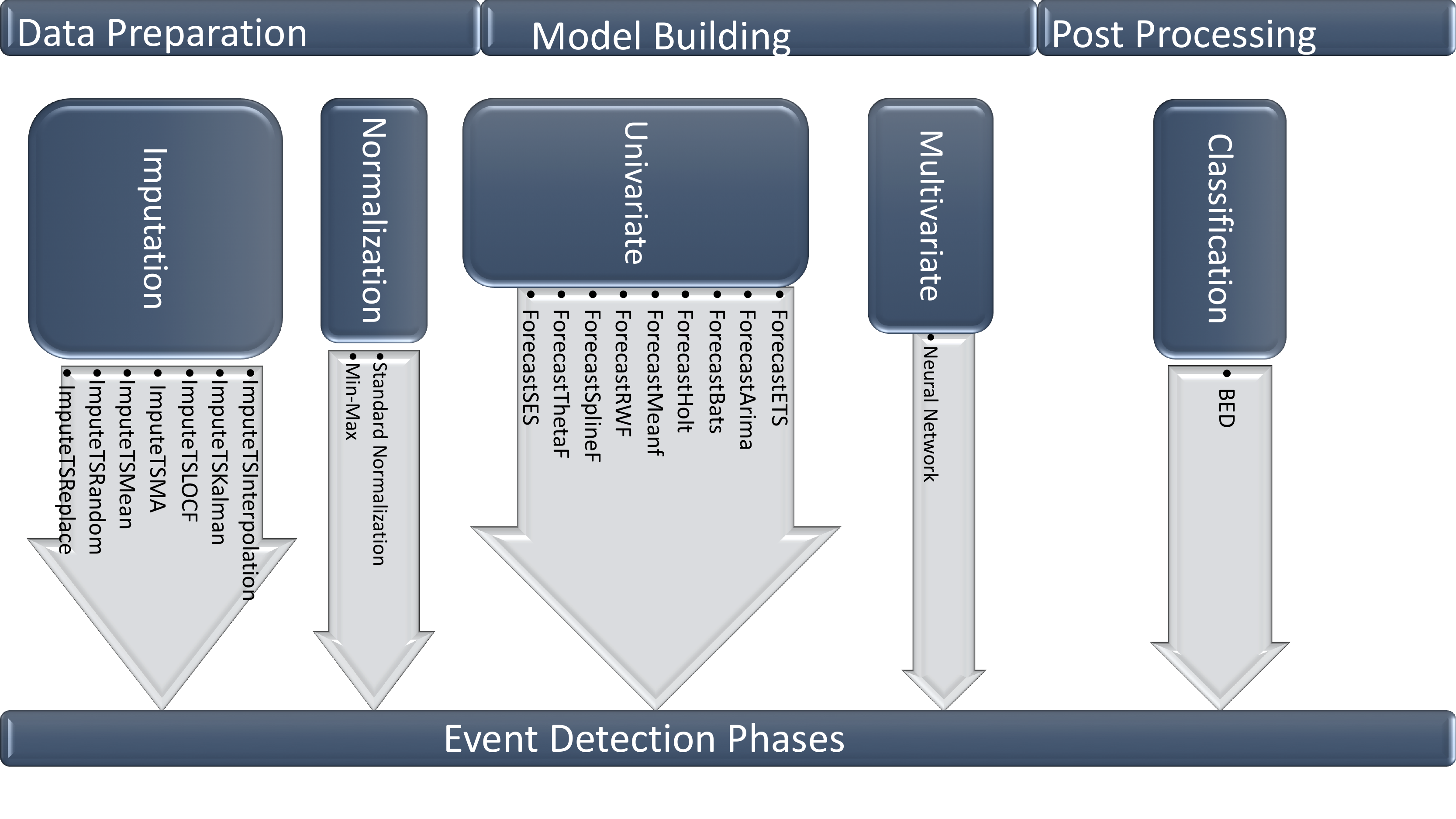}
\caption[EventDetectR-Phases]{EventDetectR-Phases}
\label{fig:e3}
\end{figure}

The event detection process is carried out as a three phase approach, namely, the data preparation phase, model building phase, and the post processing phase as shown in Figure \ref{fig:e3}. 

\subsubsection{Data Pre-Processing}

The handling of missing values is particularly an important task for event detection. If the measured values are missing, machine learning models can sometimes not be trained, or the predictions are worse. Due to various challenges in sensor transmission and  data acquisition, missing values are often found in the data.\\

In \cite{Mori15a}, the authors state that there exists no single univariate imputation technique that is suitable for all types of data patterns. Hence based on the observed data pattern, a suitable imputation technique has to be performed. To cater to this purpose, ImputeTS package \cite{Mori17a} is utilized to perform imputation. The list of available imputation techniques is in Table \ref{tab:impute}. \\

\begin{table*}[tb]
  \begin{center}
  \caption{Available Imputation Techniques}
  \begin{tabular}{lrrrrrrrr}
	\toprule
   Imputation Technique & Description\\ 
    \midrule
 	ImputeTSInterpolation& Imputation by Interpolation \\
	ImputeTSKalman & Imputation  by  Kalman  Smoothing  and  State  Space Models\\
	ImputeTSLOCF & Imputation by Last Observation Carried Forward \\
	ImputeTSMA & Imputation by Weighted Moving Average \\
	ImputeTSMean & Imputation by Mean Value\\	
	ImputeTSReplace & Imputation by a Defined Value\\
	\bottomrule
    \end{tabular}
  \label{tab:impute}
  \end{center}
\end{table*}

An additional data preparation step is data normalization. Normalization is a scaling method that is used to transform the values of various parameters into a single common scale i.e., within a specific range. The Z-score standardization is available, which transforms the raw data into a scale of zero mean and standard deviation one. If neural networks are going to be used in the modeling step, the data can be transformed with min-max normalization. Such kind of normalizing helps in preventing the network from ill-conditioning. In essence, this normalization is done to have the same range of values for each of the inputs to the model. This guarantees stable convergence of weight and biases.\\

\subsubsection{Model Building: Event Detection Algorithms}
EventDetectR supports univariate and multivariate modeling of water quality time series data. The field of univariate time series forecasting is well explored in \cite{Hynd07a}. In \cite{Hynd18a}, various forecasting algorithms are implemented and made available as an open-source R package. Considering the efficiency of these forecasting models for time series data, we intend to utilize these models to predict the quality of drinking water. Nine different forecast models are made available, as listed in Table \ref{tab:forecast}.  Based on the pattern of the time series data, appropriate forecasting models can be chosen.  

\begin{table*}[tb]
  \begin{center}
  \caption{Available Forecasting Models}
  \begin{tabular}{lrrrrrrrr}
	\toprule
   Model Name & Description\\ 
    \midrule
 	ForecastETS & Forecasting using ETS models \\
	ForecastArima & Forecasting using ARIMA or ARFIMA models \\
	ForecastBats & Forecasting using BATS and TBATS models \\
	ForecastHolt & Forecasting using Holt-Winters method \\
	ForecastMeanf & Mean Forecasting method\\	
	ForecastRWF & Naive and Random Walk Forecasting method\\
	ForecastSplineF & Cubic Spline Forecasting method\\
	ForecastThetaf & Theta forecasting method\\
	ForecastSES & Exponential smoothing forecasting method\\
	\bottomrule
    \end{tabular}
  \label{tab:forecast}
  \end{center}
\end{table*}

Also included is a multivariate neural network algorithm that models the water quality parameters through nonlinear, weighted, parametric functions. The advantage of this model is that it requires less formal statistical training. A neural network model is formulated for each of the water quality parameters using its own lagged values and the rest of all parameters involved. The error is back-propagated to the network, and the weights are adjusted back in order to reduce the error with each iteration. The neuralnet package is used for this purpose \cite{Frit16a}. 
For instance, considering 3 parameters $p1$, $p2$, $p3$, the value of $p1$ at time step $i$ is calculated as 
$p1[i] = p2[i] + p3[i] + p1[i-1]$ and the value of $p2$ at time step $i$ is calculated as 
$p2[i] = p1[i] + p3[i] + p2[i-1]$. Finally, the value of $p3$ at time step $i$ is calculated as 
$p3[i] = p1[i] + p2[i] + p3[i-1]$. 

The EventDetectR framework is developed in a generic fashion, allowing other similar models to be easily integrated.\\

\subsubsection{Post-Processing: Residual Classification}
The residuals from the model determines if each trial is a success or a failure. When residuals  are less than the  user-defined outlier threshold $\tau$, it refers to success or an inlier, whereas if residuals  are greater than the outlier threshold $\tau$, then it refers to a failure or an outlier. After this classification at each time step, the Binomial Event Discriminator (BED) is employed to obtain the continuous probability of an event ($\rho$) from the count of outliers in a specific interval \cite{Murr10a}. This specific interval is termed as a window, $BEDWindowSize$, the size of which is defined based on the user requirement. Most commonly, $BEDWindowSize$ is smaller than the sliding window $windowSize$. When $\rho$ exceeds the user-defined BED probability threshold $\vartheta$, then it signifies the occurrence of an event. \\

The binomial distribution is a discrete distribution that yields the probability of the number of success in a sequence of $n$ independent trials.\\

The probability that the water quality represents expected normal behavior in $n$ trials is represented as 
\begin{equation}
B(r;n,p)=\frac{n!}{r!(n-r)!} p^{r} q^{(n-r)},
\label{eq:bed}
\end{equation} 

where the $n$ trials is given by our BED window size $BEDWindowSize$, $q$ represents the probability that a trial succeeds and $p$ represents the probability that a trial fails as an outlier. We keep the value of both $p$ and $q$ as 0.5 and hence equation~\ref{eq:bed} is simplified to 

\begin{equation*}
B(r;n,p)=\frac{n!}{r!(n-r)!}0.5^{n}.
\label{eq:bed1}
\end{equation*}

In order to increase the probability of failure $p$ with an increase in the count of failures, the cumulative distribution function of the binomial distribution is employed as 

\begin{equation*}
P(r\leq r_{t})=\sum_{i=0}^{r_{t}}{B(r;n,p)},
\label{eq:bedcdf}
\end{equation*}

where $r_{t}$, is the probability threshold value. The advantage of this BED is that it helps in reducing the false alarm. However, as a result, there is some delay in the identification of a true event.\\

\subsection{Event Simulation Techniques}
Validating the event detection algorithms is a challenging task. In most cases, there are no data on actual events, or the occurrence of actual
events is rare. Accordingly, real test data is rare. 
Hence, an easier solution would be computer generated events. The package supports four different types of events. Namely, Sinusoidal, Ramp, Slow-sinusoidal and Square signal events. The strength and duration of the event can be controlled by the user. These artificial events are superimposed on the input data and enable the validation of the algorithms as the locations of events are known.\\

\begin{figure}[h]
\centering
\includegraphics[width=1\textwidth]{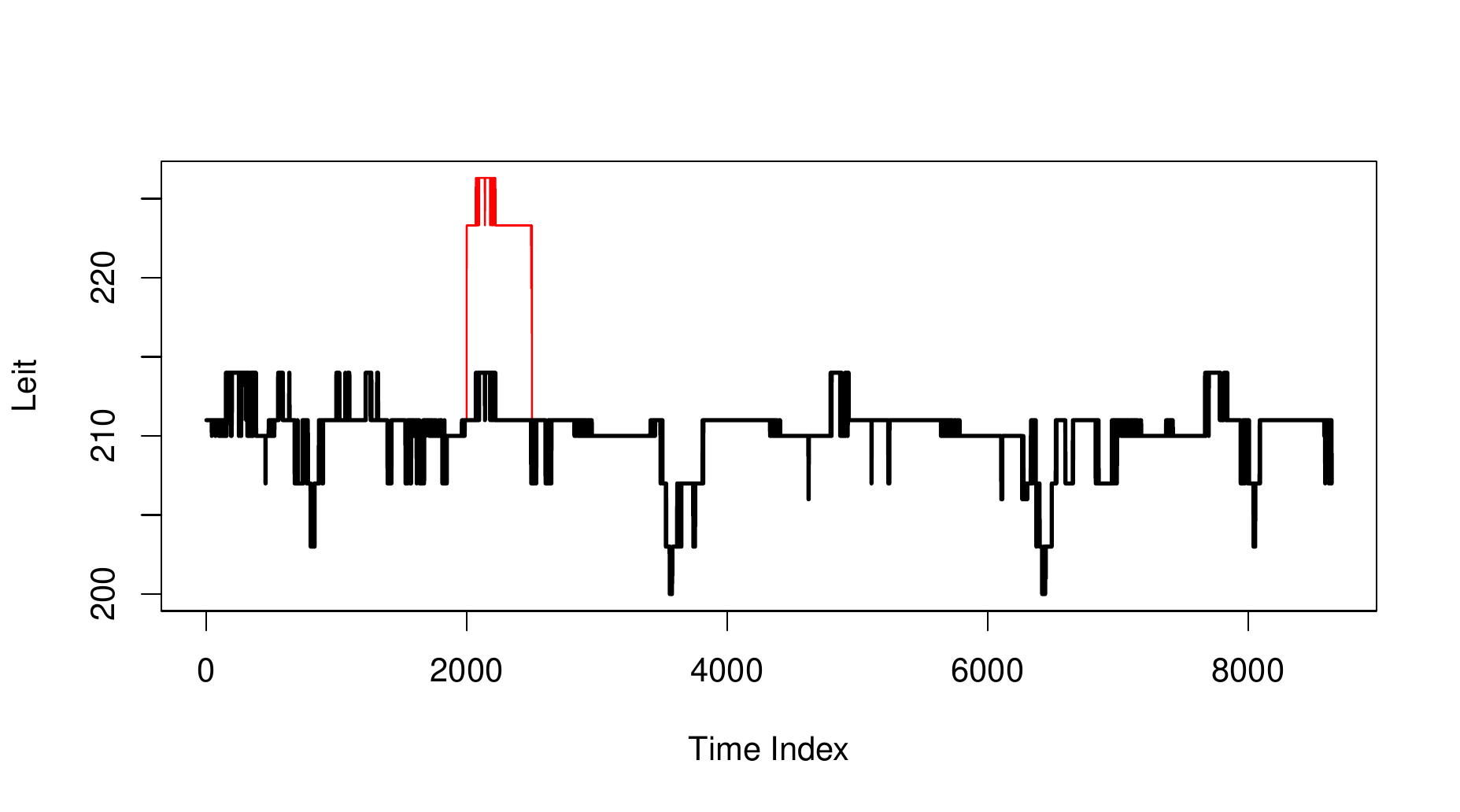}
\caption{Event Simulation on Parameter Leit from geccoIC2018Train Data-set}
\label{fig:simulate_event}
\end{figure}

The usage of the function can be illustrated using the data-set \textit{geccoIC2018Train} included on the package. An inconsistency following a square pattern can be introduced as shown in Figure \ref{fig:simulate_event}. It can be seen that from time index 2000 to 2500, the events are superimposed (showed in red). \\

\subsection{\href{https://cran.r-project.org/web/packages/EventDetectGUI/index.html}{EventDetectGUI}- R Package}
A graphical user interface (GUI) for the EventDetectR package is also available. It enables the use of the EventDetectR package for a wide range of users, mainly for practitioners without R programming skills. The user can load data into the system and visualize the data with interactive plots. After understanding the data, the user can
interactively configure the desired algorithms from EventDetectR through the GUI. The event detection results are then available to the user in a tabular fashion. Interactive graphical visualization of the results is also enabled. Additionally, the results can be exported to other formats.\\

\section{Performance Measures}\label{sec:4}
In order to evaluate the performance of the event detection algorithms, the Confusion Matrix and the Reciever Operating Curve are recommended as the most suitable performance metrics. 
\subsection{Confusion Matrix}
 The confusion matrix is formulated, as shown in Table~\ref{tab:confmatrx}. Events that are correctly detected are the true positives (TP). True events that are not detected are the false negatives (FN). Non-events that are classified as events (false alarms) represents the false positives (FP). Non-events that are not declared as events are the true negatives (TN).

\begin{table*}[tb]
  \begin{center}
  \caption{Confusion Matrix that Summarizes the Event Detection Algorithms Performance}
  \begin{tabular}{lrrrrrrrr}
	\toprule
     & Actual Event & No Actual Event\\ 
    \midrule
 	Predicted Event & TP & FP \\
	Predicted No Event & FN & TN\\
			\bottomrule
    \end{tabular}
  \label{tab:confmatrx}
  \end{center}
\end{table*}

\subsection{Receiver Operating Curve (ROC)}
\begin{figure}[h]
\centering
\includegraphics[width=1\textwidth]{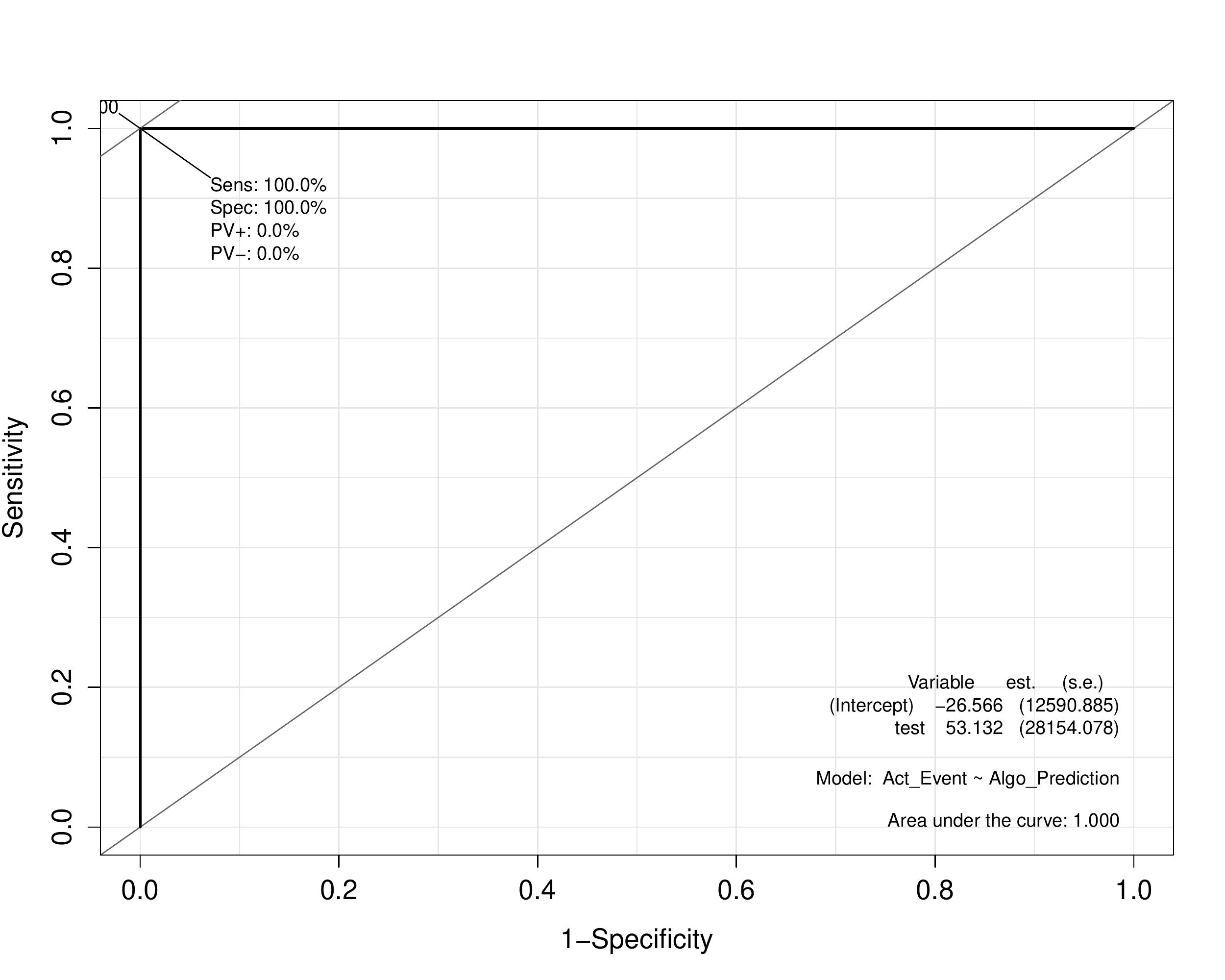}
\caption{An Ideal ROC Curve}
\label{fig:ideal_roc}
\end{figure}

A considerable trade-off is required between the numbers of TP and FP. The ROC curve is used to visualize this trade-off between
the True Positive Rate (TPR) and the False Positive Rate (FPR). For a perfect event detection algorithm, irrespective of the value of the threshold, the TPR should be equal to one, and FPR should be zero. This is shown in Figure~\ref{fig:ideal_roc}. The area under the curve, as shown in the plot is one, which is the maximum limit. The 45-degree line from the lower left to the upper right signifies the worst performing algorithm, which is the same as that of the random guess.\\

\subsection{Sensitivity and Specificity}

Sensitivity (or TPR) measures the ability of the algorithm to correctly identify an event among all the actual events. Specificity measures how accurate the event detection algorithm is against false positives or false alarms. The Accuracy, TPR rate or sensitivity, specificity, and FPR are obtained as below

\begin{align}
Accuracy=\frac{(TP+TN)}{(P+N)},\\
TPR= Sensitivity=\frac{TP}{P},\\
FPR=\frac{FP}{N},\\
Specificity= 1-FPR,\\
\label{eq:roc}
\end{align}

where P denotes the total number of events in the data-set and N denotes the total number of non-events. 
\section{Case Study}\label{sec:5}
The availability of a data-set for evaluating the performance of an event detection algorithm is very difficult. As a step to help the research community, the real data-set from the water industry is made available in the GECCO 2017 Industrial Challenge \cite{Chand17a} and the GECCO 2018 Industrial Challenge \cite{Rehb18a}. Some of these data-sets are also included as a part of the EventDetectR package.\\

\begin{figure}[h]
\centering
\includegraphics[width=1\textwidth]{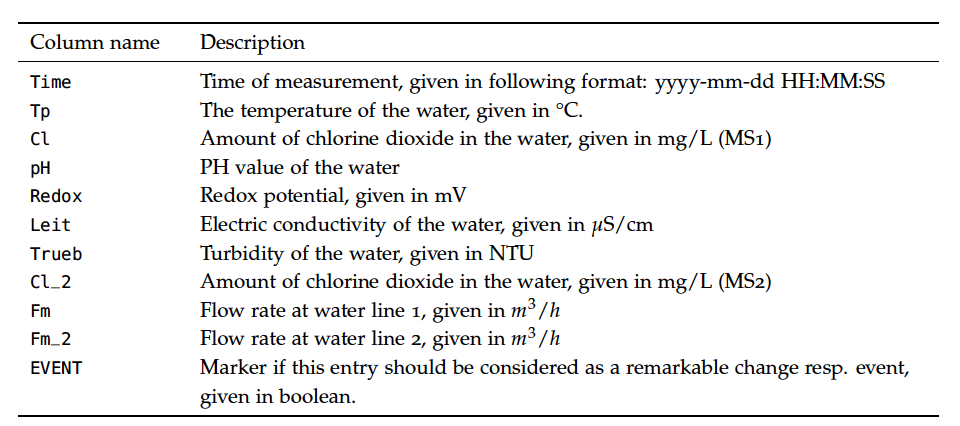}
\caption{Description of the GECCO2018 Challenge Parameters.}
\label{fig:geccodata}
\end{figure}

Let us consider the data from GECCO 2018 Industrial Challenge\cite{Rehb18a}. Figure \ref{fig:geccodata} describes the water quality data-set, which contains minutely sensor data. It consists of the amount of chlorine dioxide in the water, its pH value, the redox potential, its electric conductivity and the turbidity of the water. These are water quality parameters and any change in these is considered as an event. The flow rate and the temperature of the water are considered as operational data. Changes in these
values may indicate changes in the related quality values but are
not considered as events.

%The boxplot of the quality parameters is shown in Figure \ref{fig:bp_geccodata}. The measurements are made every minute and Figure~\ref{fig:sample_geccodata} shows an extract of the raw data.

%\begin{figure}[h]
%\centering
%\includegraphics[width=1\textwidth]{Figures/Boxplot_Geccodata.pdf}
%\caption{Boxplot of the Water Quality Parameters}
%\label{fig:bp_geccodata}
%\end{figure}

%\begin{figure}[h]
%\centering
%\includegraphics[width=1\textwidth]{Figures/Gecco_sampledata.png}
%\caption{An extract of the water quality data for a small time frame. The events are marked in red.}
%\label{fig:sample_geccodata}
%\end{figure}

For demonstration purposes, let us consider a small time frame of one week from the GECCO 2018 Industrial Challenge data-set \cite{Rehb18a}. The ForecastArima algorithm is chosen as the model. The configuration in R is done as shown below

\begin{figure}[h]
\centering
\includegraphics[width=1\textwidth]{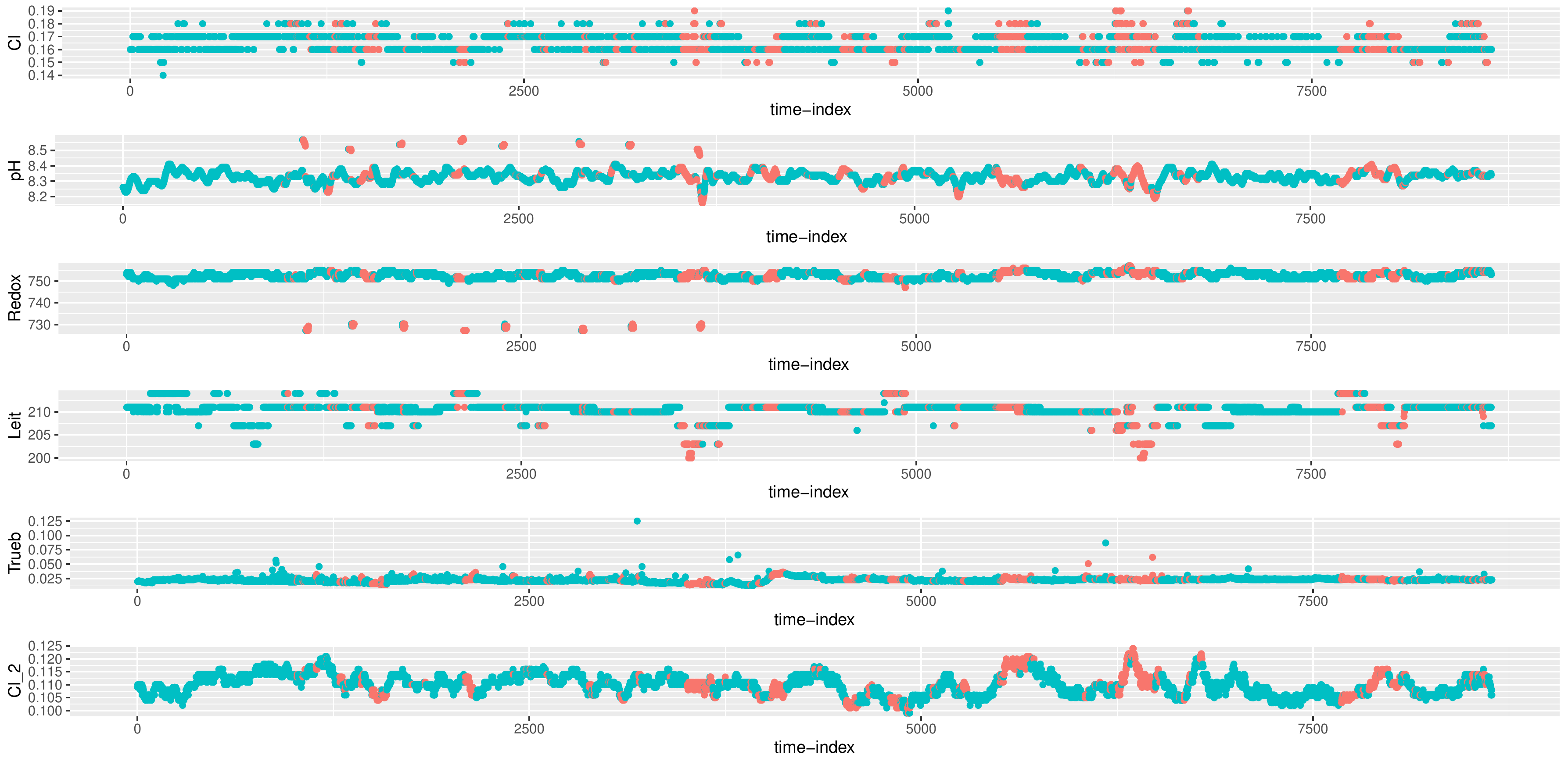}
\caption{Resulting Plot for the ForecastArima Model from EventDetectR on the Example Data-set} 
\label{fig:example_res}
\end{figure}

\begin{figure}[h]
\centering
\includegraphics[width=1\textwidth]{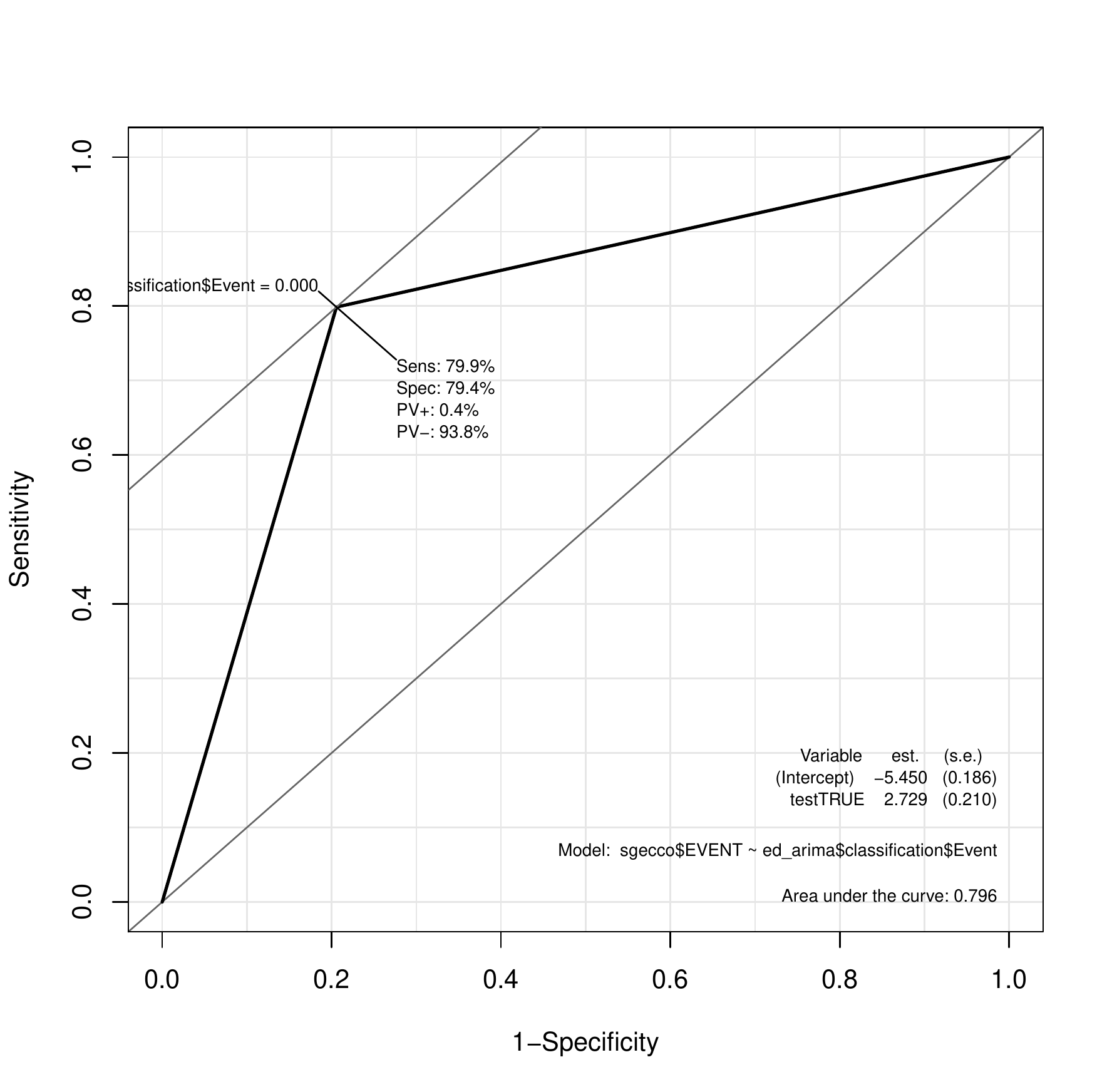}
\caption{ROC Curve for the ForecastArima Model from EventDetectR on the Example Data-set}
\label{fig:example_roc}
\end{figure}

\begin{lstlisting}[language=R]
d_int<- interval(start = as.POSIXct("2016-08-14"), 
end = as.POSIXct("2016-08-20"))
sgecco<-geccoIC2018Train[geccoIC2018Train$Time %within% d_int, ]
ed_arima <- detectEvents(gecco_data[,c(3,4,5,6,7,8)],windowSize = 1000,
 nIterationsRefit = 5,verbosityLevel = 2,
buildModelAlgo = "ForecastArima",
postProcessorControl = list(nStandardDeviationseventThreshold = 2,
eventThreshold = .7, bedWindowSize =10))
qualityStatistics(ed_arima, sgecco$EVENT)
plot(ed_arima)
\end{lstlisting}
The results obtained are shown in the form of a confusion matrix in Table \ref{tab:confmatrx_example}. Out of 144 actual events in the given time interval, 115 events were identified. There were 1748 false positives, which can be reduced with further tuning of the algorithm parameters and with the choice of other algorithms. The ROC curve is shown in Figure \ref{fig:example_roc}. The plot shows 79.9\% sensitivity and 79.4\% specificity. This indicates better performance. The resulting data plot with marked events is shown in Figure \ref{fig:example_res}. The event positions are marked for each variable in red. This plot enables further analysis to identifying the root cause of the event. 

\begin{table*}[tb]
  \begin{center}
  \caption{Confusion Matrix for the ForecastArima Model from EventDetectR on the Example Data-set}
  \begin{tabular}{lrrrrrrrr}
	\toprule
     & Actual Event & No Actual Event\\ 
    \midrule
 	Predicted Event & 115 & 1748 \\
	Predicted No Event & 29 & 6749\\
			\bottomrule
    \end{tabular}
  \label{tab:confmatrx_example}
  \end{center}
\end{table*}

\clearpage
\section{Summary and Outlook}\label{sec:6}
\href{https://cran.r-project.org/web/packages/EventDetectR/index.html}{EventDetectR} package is capable of simulating, detecting and classifying events in time-series data. It delivers an easily configurable tool for event detection. Wide range of pre-processing techniques and event detection algorithms are available. Based on the nature of the time series data, suitable pre-processing techniques and event detection algorithms can be chosen by the user. Also, the threshold limits can be modified by the user, based on their requirements. The framework is reliable with better performance and is suitable for real time event detection. The EventDetectGUI package, a graphical user interface, enables the usage of the EventDetectR package easier, especially for non-programming users.\\
\section*{Acknowledgments}

%\begin{leftbar}
This research work is funded by project Open Water Open Source(OWOS) (reference number: 005-1703-0011) and is kindly supported by the FH Zeit f{\"u}r  Forschung, Ministerium fur Innovation, Wissenschaft und Forschung des Landes NRW. \\
%\end{leftbar}
\bibliographystyle{IEEEtran}
\bibliography{Chan20a}

\begin{thebibliography}{10}
\providecommand{\url}[1]{#1}
\csname url@samestyle\endcsname
\providecommand{\newblock}{\relax}
\providecommand{\bibinfo}[2]{#2}
\providecommand{\BIBentrySTDinterwordspacing}{\spaceskip=0pt\relax}
\providecommand{\BIBentryALTinterwordstretchfactor}{4}
\providecommand{\BIBentryALTinterwordspacing}{\spaceskip=\fontdimen2\font plus
\BIBentryALTinterwordstretchfactor\fontdimen3\font minus
  \fontdimen4\font\relax}
\providecommand{\BIBforeignlanguage}[2]{{%
\expandafter\ifx\csname l@#1\endcsname\relax
\typeout{** WARNING: IEEEtran.bst: No hyphenation pattern has been}%
\typeout{** loaded for the language `#1'. Using the pattern for}%
\typeout{** the default language instead.}%
\else
\language=\csname l@#1\endcsname
\fi
#2}}
\providecommand{\BIBdecl}{\relax}
\BIBdecl

\bibitem{Stor11a}
M.~V. Storey, B.~van~der Gaag, and B.~P. Burns, ``Advances in on-line drinking
  water quality monitoring and early warning systems,'' \emph{Water Research},
  vol.~45, no.~2, pp. 741--747, 2011.

\bibitem{Ostf04a}
A.~Ostfeld and E.~Salomons, ``Optimal layout of early warning detection
  stations for water distribution systems security,'' \emph{Journal of Water
  Resources Planning and Management}, vol. 130, no.~5, pp. 377--385, 2004.

\bibitem{hach00a}
\BIBentryALTinterwordspacing
``hach.'' [Online]. Available:
  \url{https://www.hach.com/event-detection-and-security/guardianblue-early-warning-system/family?productCategoryId=35547627778}
\BIBentrySTDinterwordspacing

\bibitem{scan00a}
\BIBentryALTinterwordspacing
``s::can.'' [Online]. Available:
  \url{http://moni-tool.at/index.php/moni-tool/event-detection}
\BIBentrySTDinterwordspacing

\bibitem{Murr10a}
R.~Murray, T.~Haxton, S.~McKenna, D.~Hart, K.~Klise, M.~Koch, E.~Vugrin,
  S.~Martin, M.~Wilson, V.~Cruze \emph{et~al.}, ``Water quality event detection
  systems for drinking water contamination warning systems—development,
  testing, and application of canary,'' \emph{EPAI600IR-lOI036, US}, 2010.

\bibitem{Hart07a}
D.~Hart, S.~A. McKenna, K.~Klise, V.~Cruz, and M.~Wilson, ``Canary: a water
  quality event detection algorithm development tool,'' in \emph{Proceedings of
  the World Environmental and Water Resources Congress, ASCE, Reston, VA},
  2007, pp. 1--9.

\bibitem{Klis06a}
K.~A. Klise and S.~A. McKenna, ``Water quality change detection: multivariate
  algorithms,'' in \emph{Defense and Security Symposium}.\hskip 1em plus 0.5em
  minus 0.4em\relax International Society for Optics and Photonics, 2006, pp.
  62\,030J--62\,030J.

\bibitem{Smet06a}
E.~Smeti, L.~Kousouris, P.~Tzoumerkas, and S.~Golfinopoulos, ``Statsitical
  process control techniques on autocorrelated turbidity data from finished
  water tank,'' in \emph{Proceedings of An International Conference on Water
  Science and Technology Integrated Management on Water Resources, AQUA}, 2006.

\bibitem{Pere12a}
L.~Perelman, J.~Arad, M.~Housh, and A.~Ostfeld, ``Event detection in water
  distribution systems from multivariate water quality time series,''
  \emph{Environmental science \& technology}, vol.~46, no.~15, pp. 8212--8219,
  2012.

\bibitem{Arad13a}
J.~Arad, M.~Housh, L.~Perelman, and A.~Ostfeld, ``A dynamic thresholds scheme
  for contaminant event detection in water distribution systems,'' \emph{Water
  research}, vol.~47, no.~5, pp. 1899--1908, 2013.

\bibitem{Hou13a}
D.~Hou, H.~He, P.~Huang, G.~Zhang, and H.~Loaiciga, ``Detection of
  water-quality contamination events based on multi-sensor fusion using an
  extented dempster--shafer method,'' \emph{Measurement Science and
  Technology}, vol.~24, no.~5, p. 055801, 2013.

\bibitem{Olik14a}
N.~Oliker and A.~Ostfeld, ``A coupled classification--evolutionary optimization
  model for contamination event detection in water distribution systems,''
  \emph{Water research}, vol.~51, pp. 234--245, 2014.

\bibitem{Liu15a}
S.~Liu, H.~Che, K.~Smith, and T.~Chang, ``A real time method of contaminant
  classification using conventional water quality sensors,'' \emph{Journal of
  environmental management}, vol. 154, pp. 13--21, 2015.

\bibitem{Kuhn15a}
C.~K{\"u}hnert, M.~Baruthio, T.~Bernard, C.~Steinmetz, and J.-M. Weber,
  ``Cloud-based event detection platform for water distribution networks using
  machine-learning algorithms,'' \emph{Procedia Engineering}, vol. 119, pp.
  901--907, 2015.

\bibitem{Puig17a}
V.~Puig, C.~Ocampo-Mart{\'\i}nez, R.~P{\'e}rez, G.~Cembrano, J.~Quevedo, and
  T.~Escobet, \emph{Real-time Monitoring and Operational Control of
  Drinking-Water Systems}.\hskip 1em plus 0.5em minus 0.4em\relax Springer,
  2017.

\bibitem{Leow17a}
A.~Leow, J.~Burkhardt, W.~E. Platten~III, B.~Zimmerman, N.~E. Brinkman,
  A.~Turner, R.~Murray, G.~Sorial, and J.~Garland, ``Application of the canary
  event detection software for real-time performance monitoring of
  decentralized water reuse systems,'' \emph{Environmental Science: Water
  Research \& Technology}, vol.~3, no.~2, pp. 224--234, 2017.

\bibitem{Bila18a}
M.~Bilal, A.~Gani, M.~Marjani, and N.~Malik, ``A study on detection and
  monitoring of water quality and flow,'' in \emph{2018 12th International
  Conference on Mathematics, Actuarial Science, Computer Science and Statistics
  (MACS)}.\hskip 1em plus 0.5em minus 0.4em\relax IEEE, 2018, pp. 1--6.

\bibitem{Muha18a}
F.~Muharemi, D.~Logof{\u{a}}tu, C.~Andersson, and F.~Leon, ``Approaches to
  building a detection model for water quality: a case study,'' in \emph{Modern
  Approaches for Intelligent Information and Database Systems}.\hskip 1em plus
  0.5em minus 0.4em\relax Springer, 2018, pp. 173--183.

\bibitem{Mori15a}
S.~Moritz, A.~Sard{\'a}, T.~Bartz-Beielstein, M.~Zaefferer, and J.~Stork,
  ``Comparison of different methods for univariate time series imputation in
  r,'' \emph{arXiv preprint arXiv:1510.03924}, 2015.

\bibitem{Mori17a}
S.~Moritz and T.~Bartz-Beielstein, ``imputets: time series missing value
  imputation in r.'' \emph{R J.}, vol.~9, no.~1, p. 207, 2017.

\bibitem{Hynd07a}
R.~J. Hyndman, Y.~Khandakar \emph{et~al.}, \emph{Automatic time series for
  forecasting: the forecast package for R}.\hskip 1em plus 0.5em minus
  0.4em\relax Monash University, Department of Econometrics and Business
  Statistics~…, 2007, no. 6/07.

\bibitem{Hynd18a}
R.~Hyndman, G.~Athanasopoulos, C.~Bergmeir, G.~Caceres, L.~Chhay,
  M.~O'Hara-Wild, F.~Petropoulos, S.~Razbash, E.~Wang, and F.~Yasmeen,
  ``\BIBforeignlanguage{English}{forecast: Forecasting functions for time
  series and linear models},'' Apr. 2018.

\bibitem{Frit16a}
S.~Fritsch, F.~Guenther, and M.~F. Guenther, ``Package ‘neuralnet’,''
  \emph{The Comprehensive R Archive Network}, 2016.

\bibitem{Chand17a}
S.~Chandrasekaran, M.~Rebolledo, M.~Friese, S.~Jörg, and T.~Bartz-Beielstein,
  ``Gecco 2017 industrial challenge: Monitoring of drinking-water quality,''
  2017.

\bibitem{Rehb18a}
F.~Rehbach, S.~Moritz, S.~Chandrasekaran, M.~Rebolledo, M.~Friese, and
  T.~Bartz-Beielstein, ``Gecco 2018 industrial challenge: Monitoring of
  drinking-water quality,'' 2018.

\end{thebibliography}

\end{document}